\documentclass[final,5p,
twocolumn,nubmer]{elsarticle}

\biboptions{square,sort&compress}

\usepackage{hyperref}
\hypersetup{
    colorlinks=true,
    linkcolor=red,
    filecolor=magenta,      
    urlcolor=blue,
    citecolor = teal,
}

\usepackage{amssymb}
\usepackage{lipsum}
\usepackage{amsmath}

\usepackage{color}
\usepackage{xcolor}
\usepackage{graphicx}
\newcommand{\del}[1]{ {\color{violet}{\ifmmode\text{\sout{\ensuremath{#1}}}\else\sout{#1}\fi}} }

\definecolor{limegreen}{RGB}{154,205,0}
\definecolor{brickred}{rgb}{0.8, 0.25, 0.33}
\definecolor{amethyst}{rgb}{0.6, 0.4, 0.8}
\definecolor{azure}{rgb}{0.0, 0.5, 1.0}
\definecolor{awesome}{rgb}{1.0, 0.13, 0.32}
\usepackage[normalem]{ulem}
\newcommand{\replace}[2]{{{\color{violet}{#1}}{\color{limegreen}{\ifmmode\text{\sout{\ensuremath{#2}}}\else\sout{#2}\fi}}}}
\newcommand{\replacea}[2]{{{\color{purple}{#1}}{\color{olive}{\ifmmode\text{\sout{\ensuremath{#2}}}\else\sout{#2}\fi}}}} 
\newcommand{\replacem}[2]{{\color{azure}{#1}}{\color{brickred}{\ \ifmmode\text{\sout{\ensuremath{#2}}}\else\sout{#2}\fi}}}

\newcommand{\avr}[1]{\left\langle #1 \right\rangle}

\newcommand{\eq}[1]{(\ref{#1})}
\newcommand{\bs}{\boldsymbol}
\newcommand{\ar}[2]{{\color[rgb]{0.8,0,0.5}{
\ifmmode\text{\sout{\ensuremath{#1}}}\else\sout{#1}\fi
}}{{\color[rgb]{0.2,0.8,0.2}{\, #2}}}}
\allowdisplaybreaks

\journal{Physics Letters B}

\begin{document}

\begin{frontmatter}



\title{New mixed inhomogeneous phase in vortical gluon plasma: \\ first-principle results from rotating SU(3) lattice gauge theory}


\author[bltp]{Victor V. Braguta}
\ead{vvbraguta@theor.jinr.ru}
\author[tours,nordita]{Maxim N. Chernodub}
\ead{maxim.chernodub@univ-tours.fr}
\author[bltp]{Artem A. Roenko}
\ead{roenko@theor.jinr.ru}
\affiliation[bltp]{organization={Bogoliubov Laboratory of Theoretical Physics, Joint Institute for Nuclear Research},
            city={Dubna},
            postcode={141980}, 
            country={Russia}}
\affiliation[tours]{organization={Institut Denis Poisson UMR 7013, Universite de Tours},
            city={Tours},
            postcode={37200}, 
            country={France}}
\affiliation[nordita]{organization={Nordita, Stockholm University},
            addressline={Roslagstullsbacken 23}, 
            city={Stockholm},
            postcode={SE-106 91}, 
            country={Sweden}}

\begin{abstract}
Using first-principle numerical simulations, we find a new spatially inhomogeneous phase in rigidly rotating $N_c = 3$ gluon plasma. This mixed phase simultaneously possesses both confining and deconfining phases in thermal equilibrium. Unexpectedly, the local critical temperature of the phase transition at the rotation axis does not depend on the angular frequency within a few percent accuracy. Even more surprisingly, an analytic continuation of our results to the domain of real angular frequencies indicates a profound breaking of the Tolman-Ehrenfest law in the vicinity of the phase transition, with the confining (deconfining) phase appearing far (near) the rotation axis. 
\end{abstract}




\end{frontmatter}




\section{Introduction} 

The experimental observation of quark\--gluon plasma (QGP) produced in non-central heavy-ion collisions has ignited intense theoretical interest in these highly vortical, strongly interacting systems. The plasma droplets are created in a state possessing a nonzero net angular momentum with the vorticity of the plasma estimated to reach the enormous value of~\cite{STAR_2017ckg} 
\begin{align}
    \Omega \simeq (9 \pm 1)\times 10^{21} \, \mathrm{s}^{-1} \sim 0.03\, \mathrm{fm}^{-1} c \simeq 7\,\mathrm{MeV}\,. 
    \label{eq_omega_RHIC}
\end{align}
The rapid rotation, which is revealed experimentally via the global polarization measurements of produced hyperons~\cite{STAR_2007ccu, STAR_2017ckg, Becattini_2020ngo}, is analytically predicted to have a significant impact on the phase structure of QGP~\cite{Chen_2015hfc, Chernodub_2016kxh, Wang_2018sur, Zhang_2020hha, Sadooghi_2021upd, Fujimoto_2021xix, Chen_2022smf, Zhao_2022uxc, Chernodub_2020qah, Mameda_2023sst, Sun_2023kuu, Satapathy_2023oym, Braga_2023qej, Eto:2023tuu, Eto:2023rzd, Jiang:2023hdr}.

It must be noted that {\it all} theoretical approaches to the QGP thermodynamics are performed under the condition of the rigid rotation of the 
system~\cite{
Ambrus_2014uqa, Chen_2015hfc, Chernodub_2016kxh, Wang_2018sur, Zhang_2020hha, Sadooghi_2021upd, Fujimoto_2021xix, Golubtsova_2021agl, Chen_2022smf, Zhao_2022uxc, Chernodub_2020qah, Mameda_2023sst, Sun_2023kuu, Satapathy_2023oym, Braga_2023qej,
Eto:2023tuu, Eto:2023rzd, Jiang:2023hdr,
Braguta_2020biu, Braguta_2021jgn, Braguta:2021ucr, Braguta_2022str, Braguta:2023yjn, Yang_2023vsw}. 
In a rigidly rotating fluid, the uniformity of vorticity implies the absence of the viscous component of the stress tensor~\cite{LL6}, indicating no internal friction and no entropy production, thereby representing a thermodynamic ground state of the system. Although this simplification is not an accurate approximation of an off-equilibrium state of the QGP produced in heavy-ion collisions, it drastically simplifies the theoretical analysis and makes predictions on the consequences of rotating QCD much easier. Despite the would-be simplicity of the rigid rotation, it reveals profound disagreement between the results coming from analytical~\cite{Chen_2015hfc, Chernodub_2016kxh, Wang_2018sur, Zhang_2020hha, Sadooghi_2021upd, Fujimoto_2021xix, Chen_2022smf, Zhao_2022uxc, Chernodub_2020qah, Mameda_2023sst, Sun_2023kuu, Satapathy_2023oym, Braga_2023qej} and numerical~\cite{Braguta_2020biu, Braguta_2021jgn, Braguta:2021ucr, Braguta_2022str, Braguta:2023yjn, Yang_2023vsw} simulations (see also \cite{Jiang:2023hdr}), implying that our understanding of the rotating QGP is far from satisfactory. 

The controversy is further augmented by two independent numerical observations in finite-temperature Yang-Mills theory, made in both rotating~\cite{Braguta:2023yjn} and static~\cite{Braguta_2023kwl} systems, which revealed that the moment of inertia of the gluon plasma takes a negative value in a range of temperatures starting from a temperature slightly below the deconfining phase transition $T_{c0} \equiv T_c(\Omega = 0)$, identified in a non-rotating system, up to the ``supervortical'' temperature $T_{s} \simeq 1.5 T_{c0}$~\cite{Braguta:2023yjn, Braguta_2023kwl, Braguta_2023qex}. This exceptional phenomenon, which has no analogs in ordinary fluids, was shown to occur due to thermal evaporation of the chromomagnetic condensate~\cite{Braguta:2023yjn}. The negative moment of inertia was argued to emerge from the negativity of a gluonic Barnett effect, which implies a negative coupling of the gluon spin polarization with vorticity~\cite{Braguta_2023tqz}.

Below, we deepen this mystery even further. We reveal, using first-principle lattice simulations, a new mixed structure of confining (hadron) and deconfining (quark-gluon plasma) phases that co-exist simultaneously in a rotating hot gluon matter. Although the existence of a confining-deconfining phase was proposed some years ago~\cite{Chernodub_2020qah} (see also the holographic study~\cite{Braga_2023qej}) and was evidenced in lattice calculations with a {\it kinematically} defined order parameter~\cite{Chernodub_2022veq} we show that the {\it dynamical} structure of the plasma is surprisingly different from all existing theoretical predictions.

\section{Rotating gluodynamics on the lattice}
Rigid rotation with a fixed angular velocity $\Omega$ can be conveniently described by setting a relevant system in curved spacetime endowed with the metric~\cite{Yamamoto_2013zwa}:
\begin{align}
g_{\mu\nu} = 
\begin{pmatrix}
1 - r^2 \Omega^2 & \Omega y & -\Omega x & 0 \\
\Omega y & -1 & 0 & 0  \\  -\Omega x & 0 & -1 & 0 \\
0 & 0 & 0 & -1
\end{pmatrix}\,,
\label{eq_metric}
\end{align}
where $r = \sqrt{x^2 + y^2}$ is the distance from the axis of rotation set to be along the $z$ direction. Consequently, a rotating gluon system is described by Yang-Mills action
\begin{align}
S = - \frac{1}{2 g^{2}} \int d^{4} x\, \sqrt{- g}\,  g^{\mu \nu} g^{\alpha \beta} F_{\mu \alpha}^{a} F_{\nu \beta}^{a}\,,
\label{eq_S}
\end{align}
where $F_{\mu\nu}^a$ is the field-strength tensor of the gluon field. 

To study this system, we follow lattice methods used in Refs.~\cite{Yamamoto_2013zwa, Braguta_2020biu, Braguta:2021ucr, Braguta_2021jgn, Braguta_2022str, Braguta:2023yjn, Braguta_2023kwl, Braguta_2023qex, Braguta_2023tqz}. 
In particular, within lattice simulations, one studies the system with the action (\ref{eq_S}) in thermodynamic equilibrium. The partition function of this system can be written as the path integral over gluon fields with the weight $e^{-S_E}$, where $S_E$ is the Euclidean lattice action of rotating Yang-Mills theory. Unfortunately, this action is a complex-valued function that leads to a sign problem and does not allow to apply lattice methods directly to this theory. To overcome this obstacle, we carry out lattice simulation with imaginary angular velocity $\Omega_I = - i\Omega$ and then analytically continue the results to real values of $\Omega$ through the identification: $\Omega_I^2 \to -\Omega^2$.

To construct the lattice form of the continuum action~\eqref{eq_S}, we discretize the terms, which are coupled with the angular velocity, following Refs.~\cite{Yamamoto_2013zwa, Braguta_2021jgn},
whereas for the remaining (non-rotating) terms, we use tree-level improved Symanzik gauge action~\cite{Curci_1983an, Luscher_1985zq}. The explicit form of the used lattice action may be found in Ref.~\cite{Braguta:2023yjn}. The simulations were performed on the lattices of size $N_t\times N_z\times N_s^2$ ($N_x = N_y = N_s$), and the rotation axis passes through the lattice site in the center of $xy$-plane. To study in detail the non-trivial dynamics of gluons induced by the rotation, we use the lattices with a large extension in transversal space directions: 
$4\times 24\times 145^2$, $5\times 30\times 181^2$ and $6\times 36\times 217^2$.

In imaginary time formalism, the temperature of the system equals the inverse period along the temporal direction on the lattice. In order to avoid the particularities in the definition of temperature in the gravitational background (see Sec.~\ref{sec:TE}), we denote by $T$ the value of temperature defined precisely on the rotation axis~\cite{Braguta_2020biu, Braguta_2021jgn}, where the metric-induced effects vanish, unless stated otherwise.

We applied open boundary conditions~\cite{Braguta_2021jgn} in the directions that are orthogonal to the rotation axis and used conventional periodic boundary conditions in the directions along the rotation axis and the Euclidean time. While the choice of the boundary conditions in transversal directions warrants some discussion, the previous study~\cite{Braguta_2021jgn} indicates that a type of spatial boundary does not produce an important impact on the bulk of the system due to strong screening. Nonetheless, to validate the correctness of this approach, we reproduced some of our results with periodic boundary conditions in $xy$-plane.

\section{Slow rotation and analytic  continuity}
We carry out simulations at imaginary angular frequencies and then perform analytic  continuation to real rotations. It is convenient to describe the results in terms of the imaginary velocity $v_I = \Omega_I R$ at the QGP boundary, $v_I^2 = - v_R^2$, where the velocity $v_I$ is taken for the middle of the boundary side, i.e., at the distance $R = a (N_s - 1)/2$ from the axis of rotation. The analytic  continuation is only possible if at any point of the system $|\Omega_I| r < 1$ so that the real solid rotation does not violate causality with a corresponding theorem proven in Ref.~\cite{Ambrus_2023bid}. In our paper, we respect the causality bound.

Another important and, perhaps, not intuitively clear statement is that the thermodynamical properties of the rotating plasma, which has the shape of a cylinder, depend on the angular frequency $\Omega$ only in the combination $v_R = \Omega R$ which is nothing but the (imaginary) velocity of the system at the boundary $v_R$~\cite{Braguta:2023yjn, Braguta_2023kwl, Braguta_2023tqz}. 
In the case of slow rotation, this statement follows directly from the extensive properties of the moment of inertia, and it was numerically confirmed in lattice gluodynamics in Refs.~\cite{Braguta_2023kwl, Braguta:2023yjn} using two independent methods of calculations. For a rotating system of free bosons, it was also proven analytically in Ref.~\cite{Ambrus_2023bid}.

\section{Confining and deconfining phases}
In this paper, we concentrate on a pure gluon SU(3) system, which constitutes the most important component of the QGP and determines all its nontrivial properties. At vanishing rotation, thermal gluon matter emerges in two phases: it resides either in the confinement phase, in which all color degrees of freedom are bounded into colorless states, or the deconfinement phase, in which the asymptotic physical states are colored objects. These pure-gluon phases are precursors of the low-temperature hadronic and high-temperature QGP phases, respectively. 

Confining and deconfining phases can be distinguished with the help of the expectation of the Polyakov loop,
\begin{align}
    L({\bs r}) = {\mathrm {Tr}}\, {\mathcal P} \exp \biggl(\oint_{0}^{1/T} d \tau A_4( {\tau},{\bs r}) \biggr)\,,
    \label{eq_L}
\end{align}
where ${\mathcal P}$ is the path-ordering operator and the integral goes over the timelike component $A_4$ of the matrix-valued gluon field $A_\mu$ along the compactified Euclidean time $\tau$.

\begin{figure*}[htb]
    \includegraphics[width = 0.99\linewidth]{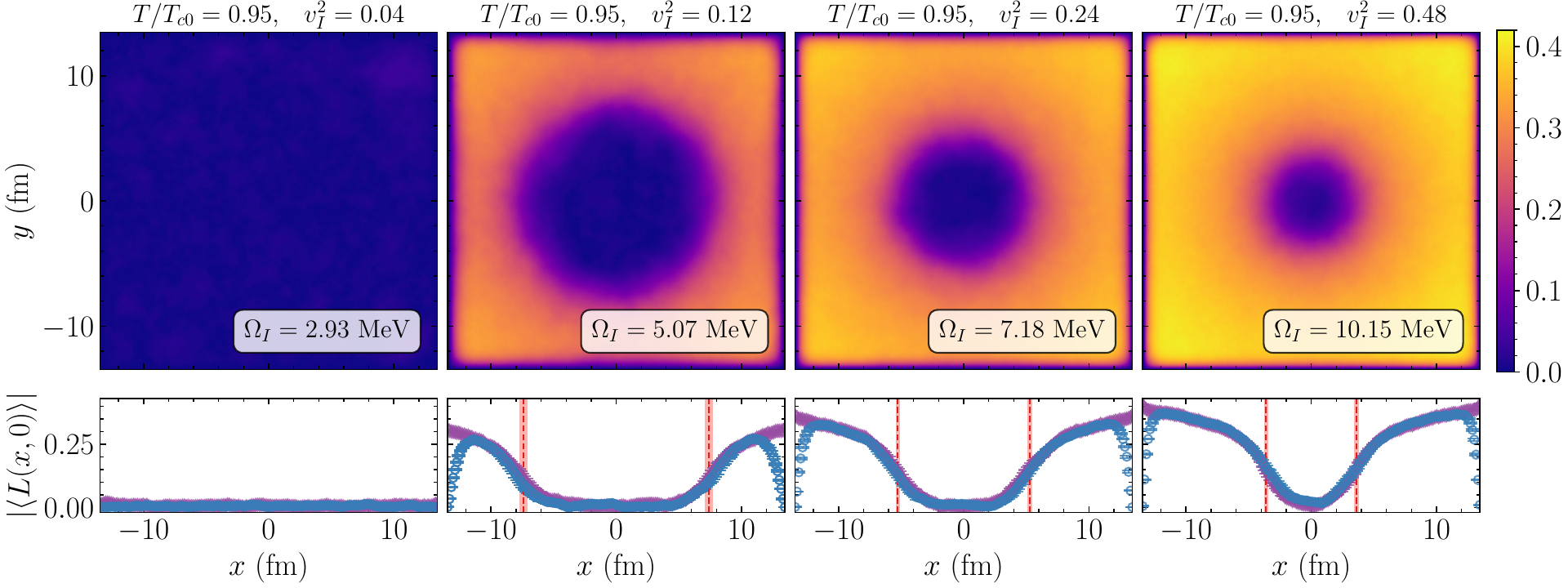}
    \caption{(top) The distribution of the local Polyakov loop in $x,y$-plane for lattice of size $5\times 30\times 181^2$ with open boundary conditions at the fixed on-axis temperature $T = 0.95\, T_{c0}$ and different imaginary angular frequencies (also shown as imaginary velocities at the boundary,  $v_I^2 \equiv (\Omega_I R)^2 = 0.04, 0.12, 0.24, 0.48$) with $R=13.5$~fm.
    (bottom) The Polyakov loop at the $x$ axis. The vertical lines mark the phase boundaries with shaded uncertainties.
    The violet (blue) data points correspond to periodic (open) boundary conditions.
    Movies on the phase evolution with increasing $\Omega_I$ are available as ancillary files~\cite{ref_SM}.
    }
    \label{fig_imshowPLxy-096}
\end{figure*}

The expectation value of the Polyakov loop operator~\eq{eq_L}, $\avr{L} = e^{- F_Q/T}$, is interpreted via the free energy $F_Q$ of an infinitely heavy test quark $Q$. In the confinement phase, the free quarks do not exist as they possess infinite energy, implying that $\avr{L} = 0$. In the deconfinement phase, on the contrary, $\avr{L} \neq 0$, the quark's free energy is finite, $F_Q \neq 0$, and the quarks can exist as free states. The expectation value of the Polyakov loop serves as a reliable order parameter that distinguishes two phases in a static, non-rotating SU(3) gluon plasma.

\section{Emergence of the inhomogeneity}
Quark-gluon plasma, slightly above the deconfining phase transition, resembles more a liquid than a gas. Our experience tells us that if a liquid is rotated --think about a rotating glass of water-- then it becomes inhomogeneous due to the centrifugal force, which literally pushes the liquid outwards the axis of rotation. Therefore, we suspect that the gluon plasma develops inhomogeneity in a rotating state, and this inhomogeneity has an imprint on its phase structure, with the phases close to the axis of rotation and far from the axis of rotation being different.

In Fig.~\ref{fig_imshowPLxy-096}, we show a local structure of the Polyakov loop in the gluon plasma for a fixed temperature and various values of $\Omega_I$. The lattice data demonstrates that gluodynamics subjected to imaginary rotation generates an inhomogeneous two-phase structure in thermal equilibrium. There are three notable features of the system:
\begin{enumerate}
   \item Imaginary rotation produces the deconfinement phase outside of the rotation axis while the region near the axis stays in the confinement phase. The deconfinement region approaches the rotation axis with the increase of $\Omega_I$;
    \item The outer, deconfining region appears even if the temperature at the rotation axis, $T$, is lower than the deconfining temperature $T_{c0}$ of a non-rotating gluon matter, $T < T_{c0}$ (so that the whole non-rotating system would reside in the confining phase at this temperature);
    \item As the on-axis temperature increases, the radius of the inner confining region shrinks.
\end{enumerate}
On the contrary, if the on-axis temperature $T$ is higher than the deconfining temperature of a non-rotating system, $T> T_{c0}$, then the two-phase structure does not emerge, and the whole imaginary-rotating system resides in the deconfinement phase.

Finalizing this section, we stress that the central confining regions in Fig.~\ref{fig_imshowPLxy-096} have the form of a disk, despite the lattice having a square shape, thus signaling the expected restoration of the rotational symmetry and implying that we work in the physical domain of lattice coupling close to the continuum limit. Moreover, the boundary conditions affect the local phase structure only very near the boundary. The latter property is a result of the short-range nature of the screening, which implies that the boundary effects on the phase structure are negligible~\cite{Braguta_2021jgn}.

\section{Size of the inhomogeneity}
In order to quantitatively study the inhomogeneous phase, it is convenient to introduce the local (pseudo)critical temperature on the rotation axis $T_c(r)$ for which the system undergoes confinement/deconfinement phase transition at a distance $r$. One has a confinement phase at distances smaller than~$r$ and a deconfinement phase at distances larger than~$r$. The local (pseudo)critical temperature is associated with the position of the peak of the Polyakov loop susceptibility, $\chi_L = \avr{|L|^2} - \avr{|L|}^2$ in the parameter space.

At a fixed distance $r$ from the rotation axis, the expectation value of the Polyakov loop and its susceptibility can only be evaluated at a finite number of spatial points proportional to the lattice extension $L_z$. Since our calculations are performed at finite $L_z$, a small volume of this lattice submanifold leads to high uncertainty in the determination of the critical temperature. To reduce the associated statistical error, we calculated the mentioned quantities within a thin cylinder $(r - \delta r/2, r + \delta r/2)$. We justified our approach by demonstrating numerically that the finiteness of $\delta r$ brings only a minor systematic error to the estimation of the critical temperature~\cite{in_preparation}.

\begin{figure}[t]
\begin{center}
\includegraphics[width = 0.99\linewidth]{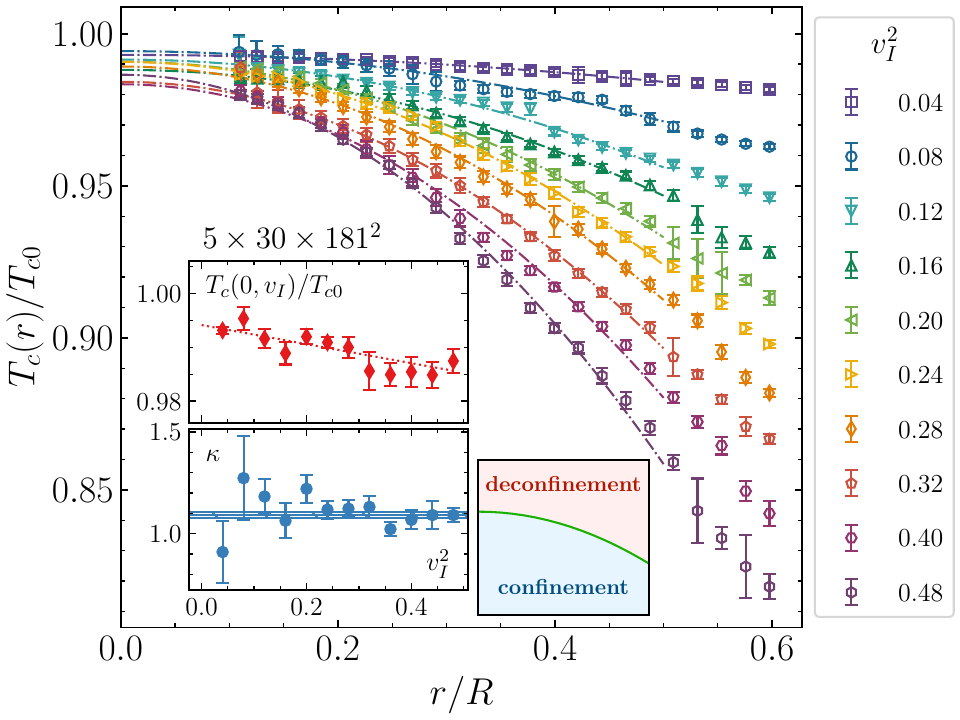}
\end{center}
\vskip -4mm 
\caption{Temperature $T_c(r)$, shown in units of the $\Omega = 0$ critical temperature $T_{c0}$ of the non-rotating system,  which should be imposed at the axis of rotation ($r=0$) in order to produce the deconfinement phase transition at the distances larger than $r$ from the rotation axis for the gluonic system rotating at various values of the imaginary angular velocity $v_I = \Omega_I R$. The dashed lines are the best quadratic fits by Eq.~\eq{eq_fit}. The insets show the arrangement of the phases for each fixed $v_I$ and the best-fit parameters vs. $v_I$. 
}
\label{fig_Tc_r}
\end{figure}

In Fig.~\ref{fig_Tc_r}, we present the local (pseudo)critical temperature $T_c(r)$ as a function of distance to the rotation axis for various imaginary angular frequencies, obtained for the averaging width $\delta r \cdot T = 3$ on the lattice with $N_t = 5$. In the absence of rotation, at $\Omega_I = 0$ (not shown in the figure), there is no dependence of the critical temperature at the center on $r$ since the transition appears simultaneously in the whole system. At any nonzero value of $\Omega_I$, the critical temperature at the rotation axis diminishes with the increase of the distance $r$ from the axis of rotation, implying that the imaginary rotation facilitates the transition to the deconfined phase outside of the rotation axis. The stronger the imaginary rotation, the lower the on-axis temperature should be to produce the deconfinement in the medium.~\footnote{The small-$r$ gap in the data presented in Fig.~\ref{fig_Tc_r} is a result of the finite thickness $\delta r$ of the cylindrical averaging manifold mentioned earlier. A decrease in $\delta r$ closes the gap and increases the statistical errors while leaving our conclusions unchanged.}

For a moderate radius $r\lesssim 0.5 R$, the critical temperature can be fitted, as a function of $r$, by the simple quadratic formula:

\begin{align}
  \frac{T_c(r,\Omega_I)} {T_{c0}} = \frac{T_c(\Omega_I)} {T_{c0}} - \kappa(\Omega_I) (\Omega_I r)^2\,,
  \label{eq_fit}
\end{align}
where the transition temperature on the rotation axis, $T_c$, and the dimensionless ``vortical curvature'' $\kappa$ serve as the fitting parameters.\footnote{The vortical curvature $\kappa$ resembles the finite-density curvature of the QCD phase transition at small values of the baryonic chemical potential~\cite{deForcrand_2002hgr}.} The best fits for various angular frequencies are shown in the main plot of Fig.~\ref{fig_Tc_r}.

\section{On-axis transition and vortical curvature} 
The results for the fit parameters are shown in the inset of Fig.~\ref{fig_Tc_r}, where the systematic uncertainties associated with the averaging width are taken into account. While both fitting parameters of Eq.~\eq{eq_fit} should, in general, depend on the imaginary frequency $\Omega_I$, our data, shown in the inset of Fig.~\ref{fig_Tc_r} as functions of the imaginary velocity at the boundary $v_I$, unexpectedly indicates that this dependence is almost absent. We believe that this tiny dependence --within a few percent of accuracy-- might be attributed to finite $N_z$ effects. 

Our result is even more surprising given that the critical temperature of the deconfining transition in {\it all} analytic  calculations available so far is predicted to exhibit a significant dependence on rotation~\cite{Chen_2015hfc, Chernodub_2016kxh, Wang_2018sur, Zhang_2020hha, Sadooghi_2021upd, Fujimoto_2021xix, Chen_2022smf, Zhao_2022uxc, Chernodub_2020qah, Mameda_2023sst, Sun_2023kuu, Satapathy_2023oym, Braga_2023qej, Jiang:2023hdr}. In addition, the previous numerical results that have found a dependence of the critical temperature on $\Omega_I$ without specifying the distance of the center of rotation~\cite{Braguta_2020biu, Braguta_2021jgn, Braguta:2021ucr, Braguta_2022str, Braguta:2023yjn, Yang_2023vsw} should be understood as the bulk-averaged results. We found a minor dependence of our results on the lattice spacing and obtained a value $\kappa = 0.902(33)$ after continuum limit extrapolation using the data for $N_t = 4,5,6$.

\section{Violation of the Tolman-Ehrenfest law} \label{sec:TE}
The temperature of a system in the thermodynamic equilibrium in an external static gravitational field depends on the coordinates $\bs r$ and obeys the well-known Tolman-Ehrenfest (TE) law~\cite{Tolman_1930zza, Tolman_1930ona}: 
$\sqrt {g_{00}({\bs r})} T({\bs r}) = T_0 = \mathrm{const}$. For a rotating system~\eq{eq_metric}, the TE law gives:
\begin{align}
    T(r) = \frac {T_0} {\sqrt{1 - \Omega^2 r^2}} = \frac {T_0} {\sqrt{1 + \Omega^2_I r^2}} \,,
    \label{eq_TE}
\end{align}
where $T_0$ is the temperature at the rotation axis ($r=0$). The last relation in Eq.~\eq{eq_TE} corresponds to the case of imaginary rotation. 
To simplify notations, we use the on-axis temperature $T_0 \equiv T$, Eq.~\eqref{eq_TE}, to refer to the temperature of the gluon plasma. 

The TE law (\ref{eq_TE}) suggests that real rotation effectively heats the system outside of the rotation axis. This fact led Ref.~\cite{Chernodub_2020qah} to conclude---also supported by a calculation in a low-dimensional confining model---that the outer region of the rotating plasma is hotter than its interior (thus, with $\avr{L} \neq 0$ far from the axis and $\avr{L} = 0$ close to the axis), which naturally leads to the two-phase structure of the rotating plasma in thermal equilibrium. For imaginary rotation, the TE law is represented by the second equality of the equation (\ref{eq_TE}), which leads to the following two-phase structure: deconfinement at the center and confinement close to the boundaries. This theoretically motivated phase structure is in contradiction with the numerical simulations of this paper. Thus, the TE law is violated in rotating gluodynamics. 

The Tolman-Ehrenfest law is violated since external gravity, generated by centrifugal forces, influences the dynamics of gluon matter and cannot be accounted for by the simple formula (\ref{eq_TE}). We suggest that this effect appears due to conformal anomaly, which is known to affect the temperature distribution in background gravitational field~\cite{Gim:2015era, Eune:2017iab}. This suggestion aligns well with the indication that a condensed-matter counterpart of the TE law, the Luttinger relation~\cite{Luttinger:1964zz}, is violated in certain materials~\cite{Park:2021oqz, Bermond_2022mjo} as a result of the conformal anomaly~\cite{Bermond_2022mjo}. Strikingly, it is the conformal anomaly, which, according to Refs.~\cite{Braguta_2023kwl, Braguta:2023yjn}, might be responsible for the negative moment of inertia of the gluon plasma~\cite{Braguta_2023tqz} so that these exotic effects can be related to each other.

\begin{figure}[t]
    \centering
    \includegraphics[width = 0.99\linewidth]{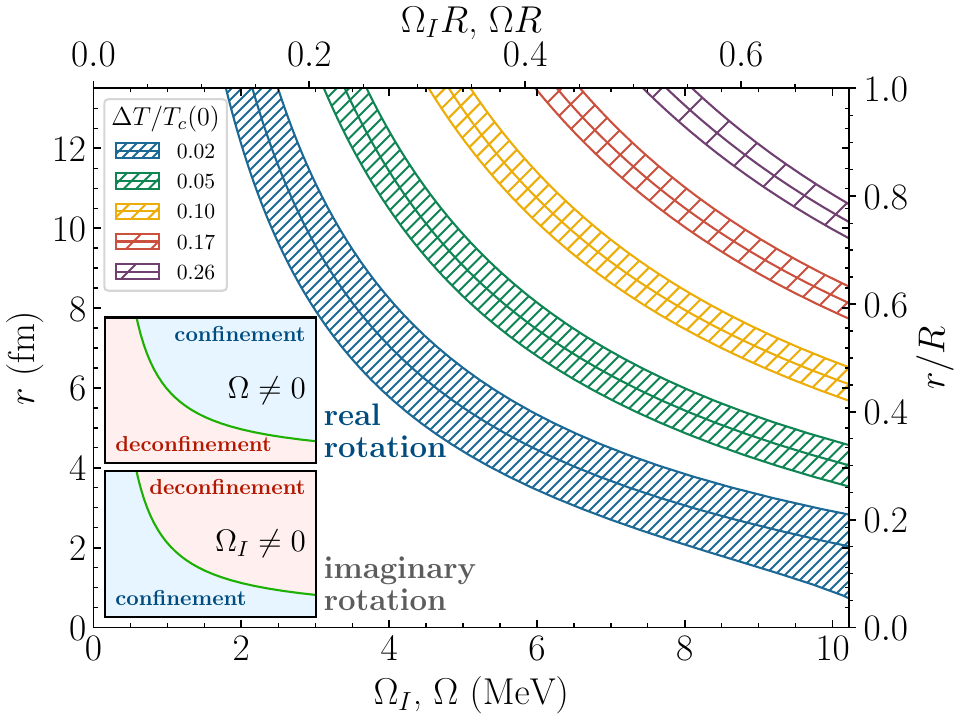}
    \caption{The critical radius $r$ of the spatial boundary between the confining and deconfining phases for imaginary frequencies $\Omega_I$ (at the axis temperature $T = T_{c0} - \Delta T$) and the real frequency $\Omega$ (at the axis temperature $T = T_{c0} + \Delta T$) in the experimentally relevant domain of parameters~\eq{eq_omega_RHIC} for various temperature offsets $\Delta T >0$. The size of the system is $R=13.5$~fm as in Fig.~\ref{fig_imshowPLxy-096}. The shading corresponds to a degree of uncertainty around the central curve. The insets show the mutual arrangement of the confining and deconfining phases for real $(\Omega)$ and imaginary $(\Omega_I)$ angular frequencies, with the details of the analytic  continuation given in Fig.~\ref{fig_analytical}.}
    \label{fig_diagram_OmegaI}
\end{figure}

\section{Phase diagram for imaginary and real rotation}
The mixed confinement-deconfinement phase may appear or not, depending on the specific values of the temperature $T$, angular velocity $\Omega$ and system radius $R$.
In order to visualize the conditions for the co-existence of phases, 
we construct the phase diagram in the $(\Omega, r)$ plane using the results for local critical temperature extrapolated to the continuum limit.\footnote{
In constructing the diagram, Fig.~\ref{fig_diagram_OmegaI}, we assumed that $T_c(0)/T_{c0} = 1$ at $v=0$ and included minor difference between $T_c(0)/T_{c0}$ in our lattice data and unity into uncertainty.
} 
The radius $r$ of the spatial boundary between the confining phase close to the axis and the deconfining phase in an outer region depends on temperature and angular frequency, as shown in Fig.~\ref{fig_diagram_OmegaI}. As the rotation increases, the radius separating these phases diminishes. This tendency is observed at all studied temperatures for imaginary rotation. The radius of the phase boundary shrinks to zero when the temperature at the center of the rotating plasma reaches the deconfining temperature of the non-rotating plasma.

\begin{figure}[t]
    \centering
    \includegraphics[width = 0.9\linewidth]{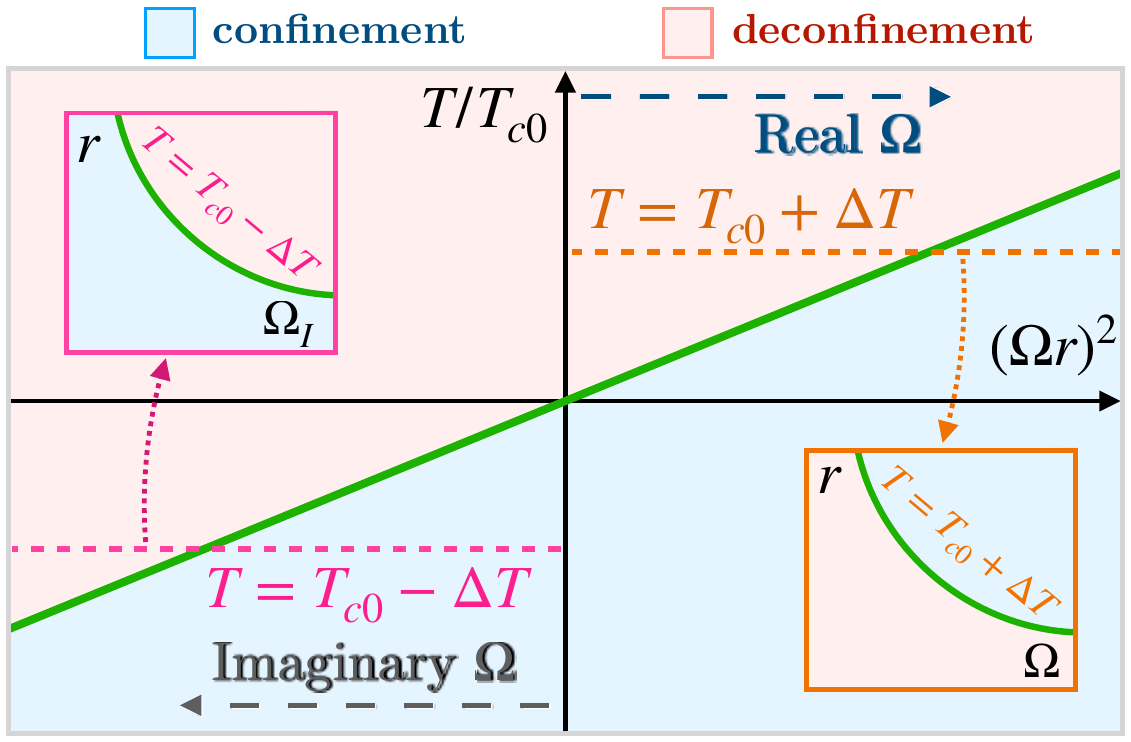}
    \caption{The analytic  continuation for slow rotation, described by Eq.~\eqref{eq_fit}, from the imaginary angular frequencies (left) to the real frequencies (right). The insets show how the phase diagrams in, respectively, $(\Omega_I,r)$ and $(\Omega,r)$ planes look like at a fixed offset  $\Delta T >0$ of the temperature $T$ of rotating plasma with respect to the $\Omega = 0$ critical temperature $T_{c0}$.}
    \label{fig_analytical}
\end{figure}

For the studied range of temperatures and angular frequencies, the phase diagram, shown in Fig.~\ref{fig_diagram_OmegaI}, is precisely the same for imaginary and real rotations up to the swap of the phases. 
In Fig.~\ref{fig_analytical} we graphically describe how the phase diagram shown in Fig.~\ref{fig_diagram_OmegaI} is built. The green straight line in Fig.~\ref{fig_analytical} represents the dependence of the local critical temperature on the local velocity of rotational motion $\Omega r$, given by Eq.~\eqref{eq_fit}. The region of imaginary rotation, $(\Omega r)^2<0$, is available for lattice simulation, whereas the results for real angular velocity, $(\Omega r)^2 > 0$, are obtained after the analytic continuation. For a given value of the temperature $T = T_{c0} - \Delta T$ the shape of the boundary between two phases on the $(\Omega_I,r)$-plane (left inset in Fig.~\ref{fig_analytical}) follows from the condition: $T_c(\Omega_I r) = T_{c0} - \Delta T$.  Graphically, this condition is represented by the intersection of the dashed line  $T=T_{c0} - \Delta T=const$ with the solid line $T_c(\Omega_I r)$.  For a symmetric value of the temperature $T = T_{c0} + \Delta T$ the critical condition is quite similar, $T_c(\Omega r) = T_{c0} + \Delta T$, but the angular velocity is real now (the resulting diagram is shown in the right inset in Fig.~\ref{fig_analytical}). The quadratic form of the critical curve~\eqref{eq_fit} ensures that the phase diagrams for these two symmetric values of the temperature $T = T_{c0} \pm \Delta T$ have the same form up to the swap of the phases and the substitution of the imaginary angular velocity to the real one. Note that this identity of the phase diagrams for real and imaginary rotation may be violated by high order $\mathcal{O}(\Omega^4)$-terms in Eq.~\eqref{eq_fit}, which are beyond the scope of this study.

\section{The role of fermions} 
In the paper~\cite{Braguta_2022str}, the influence of relativistic rotation on the QCD critical temperatures averaged over the lattice volume was studied. In particular, the authors addressed the question of how rotation influences the gluon and quark sectors of QCD. They found that rotation applied only to the gluon sector of QCD enhances the critical temperatures. On the contrary, rotation of the quark sector only decreases the critical temperatures. If one rotates both sectors simultaneously, the critical temperature increases. In addition, it was found that the quark sector of QCD plays a minor role in changing the critical temperatures due to the rotation. 
For this reason,
we expect that the observed inhomogeneous features will also persist in QGP.

\section{Summary and Conclusions}
Our results demonstrate that plasma becomes an inhomogeneous medium in which the layers of plasma situated at different distances from the axis of rotation experience the deconfining phase transition at different critical temperatures. Therefore, we conclude that there is no such notion as a single (global) critical temperature for vortical quark-gluon plasma. 

One could alternatively try to associate a rotational dependence of the critical temperature with the phase transition at the rotation axis. However, we find the absence of any noticeable dependence of the on-axis critical temperature on the angular frequency with a few percent accuracy. The latter numerical fact poses a delicate question concerning the physical sense of numerous analytical calculations based on different effective models that propose various predictions on the global effect of solid rotation on the critical temperature similar to what happens in the background magnetic field. 

We also conclude that the non-perturbative dynamics of the gluon plasma do not comply with the over-simplified picture of the Tolman-Ehrenfest (TE) law, which is based on an analogy with black body radiation. The conventional TE law --the hotter region outside and the colder region closer to the axis of rotation-- should work for a non-interacting or weakly interacting gas of particles. The outer regions of the plasma --at least, in the phenomenologically interesting phase-transition region-- turned out to be in the confinement phase, which is more appropriate for the colder environment rather than for the hot gas as prescribed by the TE picture. Being cold, the outer region does not essentially rotate.

We claim that the violation of the TE law is a consequence of the conformal anomaly in analogy with a similar effect in condensed matter systems~\cite{Bermond_2022mjo}. We suggest that the negative moment of inertia observed in our previous studies~\cite{Braguta:2023yjn, Braguta_2023kwl, Braguta_2023qex, Braguta_2023tqz} and the violation of the Tolman-Ehrenfest law found in this paper are linked to each other through a yet-to-be-discovered relationship. This link remains to be explored and is left for future work.

\section*{\bf Acknowledgements}
The work of VVB and AAR has been carried out using computing resources of the Federal collective usage center Complex for Simulation and Data Processing for Mega-science Facilities at NRC ``Kurchatov Institute'', http://ckp.nrcki.ru/ and the Supercomputer ``Govorun'' of Joint Institute for Nuclear Research. The work of VVB and AAR, which consisted of the lattice calculation of the observables used in the paper and interpretation of the data,  was supported by the Russian Science Foundation (project no. 23-12-00072). The work of MNC has been supported by the French National Agency for Research (ANR) within the project PROCURPHY ANR-23-CE30-0051-02. MNC is thankful to the members of Nordita (Stockholm) for their kind hospitality. Nordita is supported in part by Nordforsk. The authors are grateful to Dmitrii Sychev and Andrey Kotov for useful discussions. MNC thanks Frank Wilczek for valuable comments.

%
%

\bibliographystyle{elsarticle-num}
\bibliography{plasma}

\begin{thebibliography}{10}
\expandafter\ifx\csname url\endcsname\relax
  \def\url#1{\texttt{#1}}\fi
\expandafter\ifx\csname urlprefix\endcsname\relax\def\urlprefix{URL }\fi
\expandafter\ifx\csname href\endcsname\relax
  \def\href#1#2{#2} \def\path#1{#1}\fi

\bibitem{STAR_2017ckg}
L.~Adamczyk, et~al., {Global $\Lambda$ hyperon polarization in nuclear collisions: evidence for the most vortical fluid}, Nature 548 (2017) 62--65.
\newblock \href {http://arxiv.org/abs/1701.06657} {\path{arXiv:1701.06657}}, \href {https://doi.org/10.1038/nature23004} {\path{doi:10.1038/nature23004}}.

\bibitem{STAR_2007ccu}
B.~I. Abelev, et~al., {Global polarization measurement in Au+Au collisions}, Phys. Rev. C 76 (2007) 024915, [Erratum: Phys.Rev.C 95, 039906 (2017)].
\newblock \href {http://arxiv.org/abs/0705.1691} {\path{arXiv:0705.1691}}, \href {https://doi.org/10.1103/PhysRevC.76.024915} {\path{doi:10.1103/PhysRevC.76.024915}}.

\bibitem{Becattini_2020ngo}
F.~Becattini, M.~A. Lisa, {Polarization and Vorticity in the Quark\textendash{}Gluon Plasma}, Ann. Rev. Nucl. Part. Sci. 70 (2020) 395--423.
\newblock \href {http://arxiv.org/abs/2003.03640} {\path{arXiv:2003.03640}}, \href {https://doi.org/10.1146/annurev-nucl-021920-095245} {\path{doi:10.1146/annurev-nucl-021920-095245}}.

\bibitem{Chen_2015hfc}
H.-L. Chen, K.~Fukushima, X.-G. Huang, K.~Mameda, {Analogy between rotation and density for Dirac fermions in a magnetic field}, Phys. Rev. D 93~(10) (2016) 104052.
\newblock \href {http://arxiv.org/abs/1512.08974} {\path{arXiv:1512.08974}}, \href {https://doi.org/10.1103/PhysRevD.93.104052} {\path{doi:10.1103/PhysRevD.93.104052}}.

\bibitem{Chernodub_2016kxh}
M.~N. Chernodub, S.~Gongyo, {Interacting fermions in rotation: chiral symmetry restoration, moment of inertia and thermodynamics}, JHEP 01 (2017) 136.
\newblock \href {http://arxiv.org/abs/1611.02598} {\path{arXiv:1611.02598}}, \href {https://doi.org/10.1007/JHEP01(2017)136} {\path{doi:10.1007/JHEP01(2017)136}}.

\bibitem{Wang_2018sur}
X.~Wang, M.~Wei, Z.~Li, M.~Huang, {Quark matter under rotation in the NJL model with vector interaction}, Phys. Rev. D 99~(1) (2019) 016018.
\newblock \href {http://arxiv.org/abs/1808.01931} {\path{arXiv:1808.01931}}, \href {https://doi.org/10.1103/PhysRevD.99.016018} {\path{doi:10.1103/PhysRevD.99.016018}}.

\bibitem{Zhang_2020hha}
Z.~Zhang, C.~Shi, X.-T. He, X.~Luo, H.-S. Zong, {Chiral phase transition inside a rotating cylinder within the Nambu\textendash{}Jona-Lasinio model}, Phys. Rev. D 102~(11) (2020) 114023.
\newblock \href {http://arxiv.org/abs/2012.01017} {\path{arXiv:2012.01017}}, \href {https://doi.org/10.1103/PhysRevD.102.114023} {\path{doi:10.1103/PhysRevD.102.114023}}.

\bibitem{Sadooghi_2021upd}
N.~Sadooghi, S.~M.~A. Tabatabaee~Mehr, F.~Taghinavaz, {Inverse magnetorotational catalysis and the phase diagram of a rotating hot and magnetized quark matter}, Phys. Rev. D 104~(11) (2021) 116022.
\newblock \href {http://arxiv.org/abs/2108.12760} {\path{arXiv:2108.12760}}, \href {https://doi.org/10.1103/PhysRevD.104.116022} {\path{doi:10.1103/PhysRevD.104.116022}}.

\bibitem{Fujimoto_2021xix}
Y.~Fujimoto, K.~Fukushima, Y.~Hidaka, {Deconfining Phase Boundary of Rapidly Rotating Hot and Dense Matter and Analysis of Moment of Inertia}, Phys. Lett. B 816 (2021) 136184.
\newblock \href {http://arxiv.org/abs/2101.09173} {\path{arXiv:2101.09173}}, \href {https://doi.org/10.1016/j.physletb.2021.136184} {\path{doi:10.1016/j.physletb.2021.136184}}.

\bibitem{Chen_2022smf}
S.~Chen, K.~Fukushima, Y.~Shimada, {Perturbative Confinement in Thermal Yang-Mills Theories Induced by Imaginary Angular Velocity}, Phys. Rev. Lett. 129~(24) (2022) 242002.
\newblock \href {http://arxiv.org/abs/2207.12665} {\path{arXiv:2207.12665}}, \href {https://doi.org/10.1103/PhysRevLett.129.242002} {\path{doi:10.1103/PhysRevLett.129.242002}}.

\bibitem{Zhao_2022uxc}
Y.-Q. Zhao, S.~He, D.~Hou, L.~Li, Z.~Li, {Phase diagram of holographic thermal dense QCD matter with rotation}, JHEP 04 (2023) 115.
\newblock \href {http://arxiv.org/abs/2212.14662} {\path{arXiv:2212.14662}}, \href {https://doi.org/10.1007/JHEP04(2023)115} {\path{doi:10.1007/JHEP04(2023)115}}.

\bibitem{Chernodub_2020qah}
M.~N. Chernodub, {Inhomogeneous confining-deconfining phases in rotating plasmas}, Phys. Rev. D 103~(5) (2021) 054027.
\newblock \href {http://arxiv.org/abs/2012.04924} {\path{arXiv:2012.04924}}, \href {https://doi.org/10.1103/PhysRevD.103.054027} {\path{doi:10.1103/PhysRevD.103.054027}}.

\bibitem{Mameda_2023sst}
K.~Mameda, K.~Takizawa, {Deconfinement transition in the revolving bag model}, Phys. Lett. B 847 (2023) 138317.
\newblock \href {http://arxiv.org/abs/2308.07310} {\path{arXiv:2308.07310}}, \href {https://doi.org/10.1016/j.physletb.2023.138317} {\path{doi:10.1016/j.physletb.2023.138317}}.

\bibitem{Sun_2023kuu}
F.~Sun, K.~Xu, M.~Huang, {Splitting of chiral and deconfinement phase transitions induced by rotation}, Phys. Rev. D 108~(9) (2023) 096007.
\newblock \href {http://arxiv.org/abs/2307.14402} {\path{arXiv:2307.14402}}, \href {https://doi.org/10.1103/PhysRevD.108.096007} {\path{doi:10.1103/PhysRevD.108.096007}}.

\bibitem{Satapathy_2023oym}
S.~Satapathy, {Bulk viscosity of rotating, hot and dense spin 1/2 fermionic systems from correlation functions} (7 2023).
\newblock \href {http://arxiv.org/abs/2307.09953} {\path{arXiv:2307.09953}}.

\bibitem{Braga_2023qej}
N.~R.~F. Braga, O.~C. Junqueira, {Inhomogeneity of a rotating quark-gluon plasma from holography}, Phys. Lett. B 848 (2024) 138330.
\newblock \href {http://arxiv.org/abs/2306.08653} {\path{arXiv:2306.08653}}, \href {https://doi.org/10.1016/j.physletb.2023.138330} {\path{doi:10.1016/j.physletb.2023.138330}}.

\bibitem{Eto:2023tuu}
M.~Eto, K.~Nishimura, M.~Nitta, {Domain-wall Skyrmion phase in a rapidly rotating QCD matter}, JHEP 03 (2024) 019.
\newblock \href {http://arxiv.org/abs/2310.17511} {\path{arXiv:2310.17511}}, \href {https://doi.org/10.1007/JHEP03(2024)019} {\path{doi:10.1007/JHEP03(2024)019}}.

\bibitem{Eto:2023rzd}
M.~Eto, K.~Nishimura, M.~Nitta, {Non-Abelian chiral soliton lattice in rotating QCD matter: Nambu-Goldstone and excited modes}, JHEP 03 (2024) 035.
\newblock \href {http://arxiv.org/abs/2312.10927} {\path{arXiv:2312.10927}}, \href {https://doi.org/10.1007/JHEP03(2024)035} {\path{doi:10.1007/JHEP03(2024)035}}.

\bibitem{Jiang:2023hdr}
Y.~Jiang, {Rotating SU(2) gluon matter and deconfinement at finite temperature}, Phys. Lett. B 853 (2024) 138655.
\newblock \href {http://arxiv.org/abs/2312.06166} {\path{arXiv:2312.06166}}, \href {https://doi.org/10.1016/j.physletb.2024.138655} {\path{doi:10.1016/j.physletb.2024.138655}}.

\bibitem{Ambrus_2014uqa}
V.~E. Ambru\c{s}, E.~Winstanley, {Rotating quantum states}, Phys. Lett. B 734 (2014) 296--301.
\newblock \href {http://arxiv.org/abs/1401.6388} {\path{arXiv:1401.6388}}, \href {https://doi.org/10.1016/j.physletb.2014.05.031} {\path{doi:10.1016/j.physletb.2014.05.031}}.

\bibitem{Golubtsova_2021agl}
A.~A. Golubtsova, E.~Gourgoulhon, M.~K. Usova, {Heavy quarks in rotating plasma via holography}, Nucl. Phys. B 979 (2022) 115786.
\newblock \href {http://arxiv.org/abs/2107.11672} {\path{arXiv:2107.11672}}, \href {https://doi.org/10.1016/j.nuclphysb.2022.115786} {\path{doi:10.1016/j.nuclphysb.2022.115786}}.

\bibitem{Braguta_2020biu}
V.~V. Braguta, A.~Y. Kotov, D.~D. Kuznedelev, A.~A. Roenko, {Study of the Confinement/Deconfinement Phase Transition in Rotating Lattice SU(3) Gluodynamics}, JETP Lett. 112~(1) (2020) 6--12.
\newblock \href {https://doi.org/10.1134/S0021364020130044} {\path{doi:10.1134/S0021364020130044}}.

\bibitem{Braguta_2021jgn}
V.~V. Braguta, A.~Y. Kotov, D.~D. Kuznedelev, A.~A. Roenko, {Influence of relativistic rotation on the confinement-deconfinement transition in gluodynamics}, Phys. Rev. D 103~(9) (2021) 094515.
\newblock \href {http://arxiv.org/abs/2102.05084} {\path{arXiv:2102.05084}}, \href {https://doi.org/10.1103/PhysRevD.103.094515} {\path{doi:10.1103/PhysRevD.103.094515}}.

\bibitem{Braguta:2021ucr}
V.~Braguta, A.~Y. Kotov, D.~Kuznedelev, A.~Roenko, {Lattice study of the confinement/deconfinement transition in rotating gluodynamics}, PoS LATTICE2021 (2022) 125.
\newblock \href {http://arxiv.org/abs/2110.12302} {\path{arXiv:2110.12302}}, \href {https://doi.org/10.22323/1.396.0125} {\path{doi:10.22323/1.396.0125}}.

\bibitem{Braguta_2022str}
V.~V. Braguta, A.~Kotov, A.~Roenko, D.~Sychev, {Thermal phase transitions in rotating QCD with dynamical quarks}, PoS LATTICE2022 (2023) 190.
\newblock \href {http://arxiv.org/abs/2212.03224} {\path{arXiv:2212.03224}}, \href {https://doi.org/10.22323/1.430.0190} {\path{doi:10.22323/1.430.0190}}.

\bibitem{Braguta:2023yjn}
V.~V. Braguta, M.~N. Chernodub, A.~A. Roenko, D.~A. Sychev, {Negative moment of inertia and rotational instability of gluon plasma}, Phys. Lett. B 852 (2024) 138604.
\newblock \href {http://arxiv.org/abs/2303.03147} {\path{arXiv:2303.03147}}, \href {https://doi.org/10.1016/j.physletb.2024.138604} {\path{doi:10.1016/j.physletb.2024.138604}}.

\bibitem{Yang_2023vsw}
J.-C. Yang, X.-G. Huang, {QCD on Rotating Lattice with Staggered Fermions} (7 2023).
\newblock \href {http://arxiv.org/abs/2307.05755} {\path{arXiv:2307.05755}}.

\bibitem{LL6}
L.~D. Landau, E.~M. Lifshitz, Fluid Mechanics, 3rd Edition, Butterworth-Heinemann, Oxford, England, 1982.

\bibitem{Braguta_2023kwl}
V.~V. Braguta, I.~E. Kudrov, A.~A. Roenko, D.~A. Sychev, M.~N. Chernodub, {Lattice Study of the Equation of State of a Rotating Gluon Plasma}, JETP Lett. 117~(9) (2023) 639--644.
\newblock \href {https://doi.org/10.1134/S0021364023600830} {\path{doi:10.1134/S0021364023600830}}.

\bibitem{Braguta_2023qex}
V.~V. Braguta, M.~N. Chernodub, I.~E. Kudrov, A.~A. Roenko, D.~A. Sychev, {Moment of inertia and supervortical temperature of gluon plasma}, PoS LATTICE2023 (2024) 181.
\newblock \href {http://arxiv.org/abs/2311.03947} {\path{arXiv:2311.03947}}, \href {https://doi.org/10.22323/1.453.0181} {\path{doi:10.22323/1.453.0181}}.

\bibitem{Braguta_2023tqz}
V.~V. Braguta, M.~N. Chernodub, I.~E. Kudrov, A.~A. Roenko, D.~A. Sychev, {Negative Barnett effect, negative moment of inertia of (quark-)gluon plasma and thermal evaporation of chromomagnetic condensate} (10 2023).
\newblock \href {http://arxiv.org/abs/2310.16036} {\path{arXiv:2310.16036}}.

\bibitem{Chernodub_2022veq}
M.~N. Chernodub, V.~A. Goy, A.~V. Molochkov, {Inhomogeneity of a rotating gluon plasma and the Tolman-Ehrenfest law in imaginary time: Lattice results for fast imaginary rotation}, Phys. Rev. D 107~(11) (2023) 114502.
\newblock \href {http://arxiv.org/abs/2209.15534} {\path{arXiv:2209.15534}}, \href {https://doi.org/10.1103/PhysRevD.107.114502} {\path{doi:10.1103/PhysRevD.107.114502}}.

\bibitem{Yamamoto_2013zwa}
A.~Yamamoto, Y.~Hirono, {Lattice QCD in rotating frames}, Phys. Rev. Lett. 111 (2013) 081601.
\newblock \href {http://arxiv.org/abs/1303.6292} {\path{arXiv:1303.6292}}, \href {https://doi.org/10.1103/PhysRevLett.111.081601} {\path{doi:10.1103/PhysRevLett.111.081601}}.

\bibitem{Curci_1983an}
G.~Curci, P.~Menotti, G.~Paffuti, {Symanzik's Improved Lagrangian for Lattice Gauge Theory}, Phys. Lett. B 130 (1983) 205, [Erratum: Phys.Lett.B 135, 516 (1984)].
\newblock \href {https://doi.org/10.1016/0370-2693(83)91043-2} {\path{doi:10.1016/0370-2693(83)91043-2}}.

\bibitem{Luscher_1985zq}
M.~Luscher, P.~Weisz, {Computation of the Action for On-Shell Improved Lattice Gauge Theories at Weak Coupling}, Phys. Lett. B 158 (1985) 250--254.
\newblock \href {https://doi.org/10.1016/0370-2693(85)90966-9} {\path{doi:10.1016/0370-2693(85)90966-9}}.

\bibitem{Ambrus_2023bid}
V.~E. Ambru\c{s}, M.~N. Chernodub, {Rigidly rotating scalar fields: Between real divergence and imaginary fractalization}, Phys. Rev. D 108~(8) (2023) 085016.
\newblock \href {http://arxiv.org/abs/2304.05998} {\path{arXiv:2304.05998}}, \href {https://doi.org/10.1103/PhysRevD.108.085016} {\path{doi:10.1103/PhysRevD.108.085016}}.

\bibitem{ref_SM}
V.~V. Braguta, M.~N. Chernodub, A.~A. Roenko, anscillary video files, this publication (2023).

\bibitem{in_preparation}
V.~V. Braguta~et al., in preparation (2024).

\bibitem{deForcrand_2002hgr}
P.~de~Forcrand, O.~Philipsen, {The QCD phase diagram for small densities from imaginary chemical potential}, Nucl. Phys. B 642 (2002) 290--306.
\newblock \href {http://arxiv.org/abs/hep-lat/0205016} {\path{arXiv:hep-lat/0205016}}, \href {https://doi.org/10.1016/S0550-3213(02)00626-0} {\path{doi:10.1016/S0550-3213(02)00626-0}}.

\bibitem{Tolman_1930zza}
R.~C. Tolman, {On the Weight of Heat and Thermal Equilibrium in General Relativity}, Phys. Rev. 35 (1930) 904--924.
\newblock \href {https://doi.org/10.1103/PhysRev.35.904} {\path{doi:10.1103/PhysRev.35.904}}.

\bibitem{Tolman_1930ona}
R.~Tolman, P.~Ehrenfest, {Temperature Equilibrium in a Static Gravitational Field}, Phys. Rev. 36~(12) (1930) 1791--1798.
\newblock \href {https://doi.org/10.1103/PhysRev.36.1791} {\path{doi:10.1103/PhysRev.36.1791}}.

\bibitem{Gim:2015era}
Y.~Gim, W.~Kim, {A Quantal Tolman Temperature}, Eur. Phys. J. C 75~(11) (2015) 549.
\newblock \href {http://arxiv.org/abs/1508.00312} {\path{arXiv:1508.00312}}, \href {https://doi.org/10.1140/epjc/s10052-015-3765-2} {\path{doi:10.1140/epjc/s10052-015-3765-2}}.

\bibitem{Eune:2017iab}
M.~Eune, W.~Kim, {Proper temperature of the Schwarzschild AdS black hole revisited}, Phys. Lett. B 773 (2017) 57--61.
\newblock \href {http://arxiv.org/abs/1703.00589} {\path{arXiv:1703.00589}}, \href {https://doi.org/10.1016/j.physletb.2017.08.009} {\path{doi:10.1016/j.physletb.2017.08.009}}.

\bibitem{Luttinger:1964zz}
J.~M. Luttinger, {Theory of Thermal Transport Coefficients}, Phys. Rev. 135 (1964) A1505--A1514.
\newblock \href {https://doi.org/10.1103/PhysRev.135.A1505} {\path{doi:10.1103/PhysRev.135.A1505}}.

\bibitem{Park:2021oqz}
J.~Park, O.~Golan, Y.~Vinkler-Aviv, A.~Rosch, {Thermal Hall response: Violation of gravitational analogs and Einstein relations}, Phys. Rev. B 105~(20) (2022) 205419.
\newblock \href {http://arxiv.org/abs/2108.06162} {\path{arXiv:2108.06162}}, \href {https://doi.org/10.1103/PhysRevB.105.205419} {\path{doi:10.1103/PhysRevB.105.205419}}.

\bibitem{Bermond_2022mjo}
B.~Bermond, M.~Chernodub, A.~G. Grushin, D.~Carpentier, {{Anomalous Luttinger equivalence between temperature and curved spacetime: From black holes to thermal quenches}}, SciPost Phys. 16 (2024) 084.
\newblock \href {http://arxiv.org/abs/2206.08784} {\path{arXiv:2206.08784}}, \href {https://doi.org/10.21468/SciPostPhys.16.3.084} {\path{doi:10.21468/SciPostPhys.16.3.084}}.

\end{thebibliography}

\end{document}